\documentclass[conference]{IEEEtran}
\IEEEoverridecommandlockouts
\usepackage{cite}
\usepackage{overpic}
\usepackage{amsmath,amssymb,amsfonts}
\usepackage{graphicx}
\usepackage{xcolor}
\usepackage{amsthm}
\usepackage{graphicx}
\usepackage{epstopdf}
\usepackage{amsmath,bm}
\usepackage{amsfonts}
\usepackage{amssymb}
\usepackage{color}
\usepackage{multirow}
\usepackage{multicol}
\usepackage{algorithm}
\usepackage{url}

\usepackage{algorithmic}

\usepackage{caption}
\usepackage{subcaption}

\ifCLASSOPTIONcompsoc
  \usepackage[caption=false,font=normalsize,labelfont=sf,textfont=sf]{subfig}
\else
  \usepackage[caption=false,font=footnotesize]{subfig}
\fi

\theoremstyle{plain}

\newcommand{\vect}[1]{\mathbf{#1}}

\def\diag{\mathrm{diag}}

\def\Htran{\mbox{\tiny $\mathrm{H}$}}
\def\Ttran{\mbox{\tiny $\mathrm{T}$}}

\def\sinc{\mathrm{sinc}}

\begin{document}

\title{Experimental Validation of Reflective\\ Near-Field Beamfocusing using a $b$-bit RIS}

\author{\IEEEauthorblockN{ Emil Björnson, Murat Babek Salman}
\IEEEauthorblockA{\textit{Department of Communication Systems, KTH Royal Institute of Technology, Sweden}}
\IEEEauthorblockA{
 emilbjo@kth.se, mbsalman@kth.se}
\thanks{This work was supported by the Swedish Foundation for Strategic Research (FFL18-0277 grant) and the Swedish Research Council (FREEDOM grant).}%
}

\maketitle
\begin{abstract}
This paper presents the first experimental validation of reflective near-field beamfocusing using a reconfigurable intelligent surface (RIS). While beamfocusing has been theoretically established as a key feature of large-aperture RISs, its practical realization has remained unexplored. We derive new analytical expressions for the array gain achieved with a $b$-bit RIS in near-field line-of-sight scenarios, characterizing both the finite depth and angular width of the focal region. The theoretical results are validated through a series of measurements in an indoor office environment at 28 GHz using a one-bit 1024-element RIS. The experiments confirm that beamfocusing can be dynamically achieved and accurately predicted by the proposed simplified analytical model, despite the presence of hardware imperfections and multipath propagation. These findings demonstrate that near-field beamfocusing is a robust and practically viable feature of RIS-assisted wireless communications.
\end{abstract}

\vspace{-2mm}

\section{Introduction}

Reconfigurable intelligent surfaces (RISs) have emerged as a new kind of repeater that through their reflections can simultaneously amplify signals in desired directions and change the shape of wavefronts \cite{Bjornson2022a}. An RIS consists of a two-dimensional array of subwavelength-sized elements with reconfigurable reflection coefficients, which can be tuned to modify how the incident waves are scattered.
The previous literature contains plenty of algorithms for selecting the RIS configuration to 
enhance communication capacity-related metrics by shaping the propagation environment \cite{Pan2022a}.
Several key RIS features have been validated experimentally, including the cascaded pathloss formula \cite{Tang2021a}, over-the-air RIS configuration in outdoor scenarios \cite{Pei2021a}, the angular beampatterns \cite{Trichopoulos2020a,Rossanese2024a}, and how an RIS can encode information by switching its configuration \cite{Shao2025}.

Many RIS configuration algorithms identify desirable phase-shifts in the full range $[-\pi,\pi]$, but practical RIS elements can usually only implement values from a discrete set $\mathcal{Q}$ \cite{Wu2020b}. A simple way to handle this is to quantize the desired phases to the closest points in $\mathcal{Q}$. The resulting performance losses with Rayleigh fading channels are quantified in \cite{Wu2020b,Enqvist2024a}, among others.
More refined algorithms have been developed and experimentally validated in \cite{Pei2021a, Trichopoulos2020a, Sanchez2021a, Sang2024a}, with a focus on the signal-to-noise ratio (SNR) achieved at the receiver.

While much of the early research on RIS was developed under plane-wave far-field assumptions, many practical deployments, especially at millimeter-wave and terahertz frequencies with electrically large apertures and short link distances, operate in the radiative near-field (Fresnel) regime. 
The boundary between these regimes is commonly characterized by the Fraunhofer distance, $d_{\textrm{F}} = 2D^2/\lambda $, where $D$ is the aperture length and $\lambda$ is the wavelength \cite{Bjornson2021b}.
Wireless propagation in the radiative near-field is characterized by spherical wavefronts, and it has been theoretically shown that an RIS can control these wavefronts to focus reflected signals on a focal point and achieve a finite beamdepth around it.
At 28 GHz, a relatively small RIS with an array size of $17\times 17$\,cm has a Fraunhofer distance of $10.8$\,m, so the near-field covers an entire room of an indoor network.
Finite-depth beamfocusing enables an RIS to amplify the SNR at the intended receiver while reducing interference at unintended places \cite{Mu2024}.
It has been analyzed theoretically in \cite{Bjornson2021b,Bjornson2020a, Jiang2023a,Kolomvakis2025a,Yu2025a} among others.
To the best of our knowledge, the only experimental validation is \cite{Mei2022a}, where three non-reconfigurable metasurfaces with different focal distances were manufactured and utilized to showcase beamfocusing in an anechoic chamber.
Hence, beamfocusing with an RIS remains to be demonstrated experimentally.

In this paper, we address this gap through a dedicated experimental campaign in a real indoor environment. In particular, we study finite-depth beamfocusing with a $b$-bit RIS by first deriving novel theoretical array gain expressions and then comparing them to measurement results. 
We demonstrate that beamfocusing is not a theoretical curiosity, but can be achieved in the studied office scenarios, that a limited RIS resolution does not widen the depth of focus, and that simple models are sufficient to predict the behavior of practical hardware.

\textit{Outline:} The basic system model with ideal hardware is provided in Section~\ref{sec:system-model}.
We derive novel theoretical formulas for the depth and width of reflective beamfocusing with $b$-bit RIS in Section~\ref{sec:theoretical_formulas}. These formulas are validated experimentally in Section~\ref{sec:experiments} and the conclusions are summarized in Section~\ref{sec:conclusion}.

\section{System Model and Ideal RIS Configuration}
\label{sec:system-model}

In this paper, we consider a communication link between a single-antenna transmitter and a single-antenna receiver, assisted by an RIS equipped with $N$ reflecting elements. The received signal $r \in \mathbb{C}$ can be expressed as \cite{Bjornson2022a}
\begin{equation}
    r = \underbrace{\left(  h_{\textrm{s}} + \vect{h}_{\textrm{r}}^{\Ttran} \vect{\Phi} \vect{h}_{\textrm{t}} \right)}_{=h} s + n,
\end{equation}
where $s$ is the transmitted signal and $n$ is the receiver noise.
The end-to-end channel coefficient $h \in \mathbb{C}$ consists of two components.
The static paths that does not involve the RIS are jointly represented by $h_{\textrm{s}}$, while the reconfigurable path via the RIS is $\vect{h}_{\textrm{r}}^{\Ttran} \vect{\Phi} \vect{h}_{\textrm{t}}$, where $\vect{h}_{\textrm{t}} = [h_{\textrm{t},1}, \ldots, h_{\textrm{t},N}]^{\Ttran} \in \mathbb{C}^N$ is the channel from the transmitter to the RIS, $\vect{h}_{\textrm{r}} = [h_{\textrm{r},1}, \ldots, h_{\textrm{r},N}]^{\Ttran} \in \mathbb{C}^N$ is the channel from the RIS to the receiver, and  $\boldsymbol{\Phi} = \mathrm{diag}\left(e^{j\phi_{1}}, \ldots, e^{j\phi_{N}}\right)$ is the diagonal matrix containing the $N$ reconfigurable phase-shifts in the RIS. Note that this model neglects mutual coupling and nonlinearities, and we will later show that it is anyway sufficiently accurate when studying practical RIS that are configured to mitigate these effects.

A reasonable goal for this system is to configure the RIS to maximize the channel gain $|h|^2$, which is achieved by \cite{Bjornson2022a}
\begin{equation} \label{eq:phi_n}
\phi_n = \arg(h_{\textrm{s}}) - \arg(h_{\textrm{r},n}h_{\textrm{t},n}) \quad \mathrm{mod} \,\, 2 \pi.
\end{equation}
This configuration not only maximizes the SNR at the receiver, but also shapes the wave reflected by the RIS. This is particularly evident when there is a line-of-sight (LOS) channel from the RIS to the receiver, in which case it will create an angular beam aimed toward the receiver, if it is located in the far-field of the RIS.
In this paper, we are particularly interested in the behaviors in the radiative near-field, for which theoretical formulas indicate that the beam will also have finite depth~\cite{Bjornson2021b}.

Many practical RIS elements only support a finite number of phase-shift values, and this also applies to the RIS used for experimental validation in Section~\ref{sec:experiments}.
Particularly, a $b$-bit RIS supports $2^b$ phase-shifts uniformly distributed on the unit circle. If the optimal value in \eqref{eq:phi_n} is quantized to the nearest implementable point, the phase deviation lies in the set $\mathcal{Q}_b=[-\frac{\pi}{2^b},\frac{\pi}{2^b}]$.
We evaluate its negative impact in the next section.

\section{Near-Field Beamfocusing with a $b$-bit RIS}
\label{sec:theoretical_formulas}

In this section, we will analyze the array gains achieved when using a $b$-bit RIS in a near-field LOS scenario. We are interested in how the array gain varies around the focal point of the reflection, which characterizes its depth and width.

Suppose the RIS is centered around the origin in the YZ-plane and the $n$th RIS element is located at $[0, y_n, z_n]^{\Ttran}$. The transmitter is located at the distance $d_{\textrm{t}}$
in the azimuth angle direction $\varphi_{\textrm{t}}$ and elevation angle direction $\theta_{\textrm{t}}$. The corresponding parameters for the receiver are the distance $d_{\textrm{r}}$, the azimuth angle $\varphi$, and elevation angle $\theta_{\textrm{r}}$.
It follows that $\vect{h}_{\textrm{t}} = \alpha_{\textrm{t}} \vect{a}(\varphi_{\textrm{t}},\theta_{\textrm{t}},d_{\textrm{t}})$ and
$\vect{h}_{\textrm{r}} = \alpha_{\textrm{r}} \vect{a}(\varphi_{\textrm{r}},\theta_{\textrm{r}},d_{\textrm{r}})$, 
 where $\alpha_{\textrm{t}} \in \mathbb{C}$ and $\alpha_{\textrm{r}} \in \mathbb{C}$ are the channel coefficients from the origin to the transmitter and receiver, respectively, and the near-field array response vector is defined as \cite{Kosasih2025a}
\begin{equation}
    \vect{a}(\varphi,\theta,d) = \left[ e^{-j \frac{2\pi}{\lambda} (d_1-d)},\ldots, e^{-j \frac{2\pi}{\lambda} (d_N-d)} \right]^{\Ttran},
\end{equation}
where $d_n$ is the distance from element $n$ to the point $(\varphi,\theta,d)$:
\begin{align} \nonumber
    &d_n = \left\| \begin{bmatrix}
        d \cos(\varphi) \cos(\theta) \\
        d \sin(\varphi) \cos(\theta) \\
        d \sin(\theta)
    \end{bmatrix} - \begin{bmatrix}
        0 \\
        y_n \\
        z_n
    \end{bmatrix} \right\| \\ 
    &= d \sqrt{1 - \frac{2(\sin(\varphi) \cos(\theta) y_n +  \sin(\theta) z_n)}{d} + \frac{y_n^2+z_n^2}{d^2}}. \label{eq:d_n}
\end{align}
Assuming a negligible static path for notational convenience (i.e., $h_{\textrm{s}}=0$), a $b$-bit RIS configuration based on \eqref{eq:phi_n} becomes
\begin{align} \label{eq:phi_n_errors}
\vect{\Phi} = \diag \left(\vect{a}^{\Htran}(\varphi_{\textrm{t}},\theta_{\textrm{t}},d_{\textrm{t}}) \odot \vect{a}^{\Htran}(\varphi_{\textrm{r}},\theta_{\textrm{r}},d_{\textrm{r}}) \right) \diag (e^{j \epsilon_1},\ldots,e^{j \epsilon_N}),
\end{align}
where $\epsilon_n \in \mathcal{Q}_b$ is the phase-deviation for element $n$, $\diag(\cdot)$ creates a diagonal matrix, and $\odot$ denotes the Hadamard product.
With this configuration,  the channel gain becomes
\begin{align} 
    |h|^2 = | \vect{h}_{\textrm{r}}^{\Ttran} \vect{\Phi} \vect{h}_{\textrm{t}}|^2 = \left| \alpha_{\textrm{r}} \alpha_{\textrm{t}} \sum_{n=1}^{N} e^{j \epsilon_n} \right|^2.
\end{align}
It would equal $N^2 | \alpha_{\textrm{r}} \alpha_{\textrm{t}} |^2$ if $\epsilon_1=\ldots=\epsilon_N=0$.
However, these phase deviations are non-zero when using a practical $b$-bit RIS.
The deviations are deterministic for LOS channels and can be computed for a specific RIS setup; however, since there are many parameters, their joint effect appears random-like. Hence, to quantify their impact, we approximate them as independent random variables that are
uniformly distributed in $\mathcal{Q}_b$, just as in the independent Rayleigh fading scenario considered in \cite{Wu2020b,Enqvist2024a}. The average channel gain then becomes
\begin{align} \nonumber
    \mathbb{E} \{ |h|^2 \} &=  \left| \alpha_{\textrm{r}} \alpha_{\textrm{t}} \right|^2
\sum_{n=1}^{N} \sum_{m=1}^{N} \mathbb{E} \{ e^{j \epsilon_n} e^{-j \epsilon_m} \}  \\
& = \left| \alpha_{\textrm{r}} \alpha_{\textrm{t}} \right|^2 \left( N^2 \sinc^2 (2^{-b}) + N \left(1-\sinc^2(2^{-b}) \right) \right), \label{eq:h_gain}
\end{align}
by utilizing the fact that $\mathbb{E} \{ e^{j \epsilon_n} \} = \sinc(2^{-b})$ \cite[Eq.~(22)]{Enqvist2024a}. This channel gain expression differs from those with Rayleigh fading in \cite{Wu2020b,Enqvist2024a}, but shares the property of being proportional to $N^2$ (the array gain) and $\sinc^2(2^{-b}) \in [0.4,1]$. The latter is the loss due to finite resolution and approaches $1$ as $b\to \infty$.

\begin{figure} 
        \centering 
	\begin{overpic}[width=\columnwidth,tics=10]{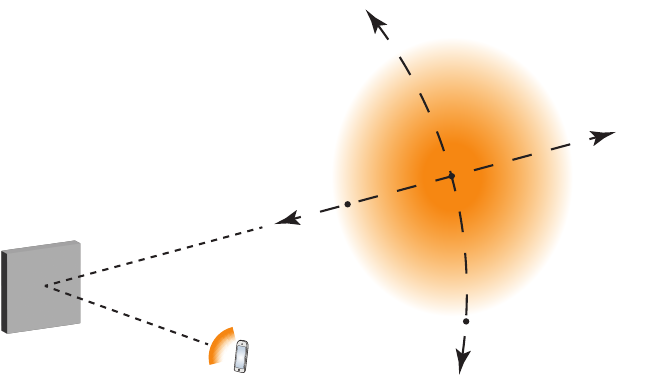}
  \put(0.5,23){\small Reflective RIS}
  \put(39,1){\small Transmitter}
    \put(38,28){\small $(\varphi_{\textrm{r}},\theta_{\textrm{r}},F)$}
  \put(72,9){\small $(\varphi_{\textrm{F}},\theta_{\textrm{F}},d_{\textrm{r}})$}
   \put(69,33){\small \rotatebox{13}{Receiver}}
   \put(70,28){\small \rotatebox{13}{$(\varphi_{\textrm{r}},\theta_{\textrm{r}},d_{\textrm{r}})$}}
\end{overpic} \vspace{-5mm}
        \caption{The receiver is located at $(\varphi_{\textrm{r}},\theta_{\textrm{r}},d_{\textrm{r}})$. We explore how the channel gain changes when the focal point of the RIS reflection is mismatched in either distance or angle.}
\label{focusing_example} 
\end{figure}
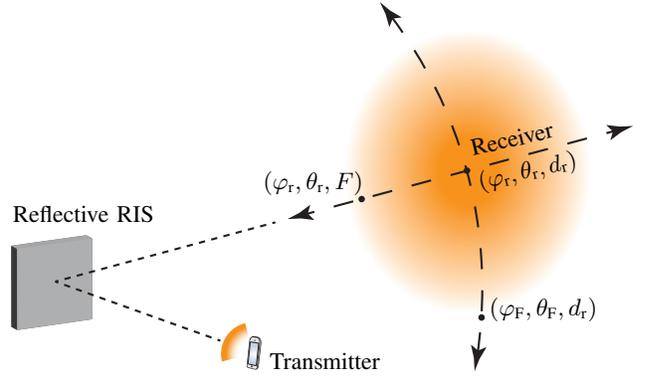

\subsection{Depth of Beamfocusing}

We will now characterize the shape of the beam reflected from the RIS, especially in the radiative near-field, where the depth can be finite. Suppose the RIS is configured to focus its reflected signal on a point in the correct direction, but at a different focal distance $F$ as illustrated in Fig.~\ref{focusing_example}. We want to quantify the resulting loss in channel gain analytically.

To this end, suppose the RIS configuration is selected as in \eqref{eq:phi_n_errors} but with $d_{\textrm{r}}$ replaced by $F$, so the channel gain becomes
\begin{align} 
    |h|^2 = |\alpha_{\textrm{r}} \alpha_{\textrm{t}} |^2
    \underbrace{\left| \vect{a}^{\Htran}(\varphi_{\textrm{r}},\theta_{\textrm{r}},F) \diag (e^{j \epsilon_1},\ldots,e^{j \epsilon_N} ) \vect{a}(\varphi_{\textrm{r}},\theta_{\textrm{r}},d_{\textrm{r}}) \right|^2}_{\triangleq G(F,d_{\textrm{r}})}, \label{eq:h2_beamfocusing}
\end{align}
where we call $G(F,d_{\textrm{r}})$ the \emph{array gain function}. To evaluate it, we consider the first-order Taylor approximation of \eqref{eq:d_n}:
\begin{align} \label{eq:Taylor_approx}
    d_n \approx d - (\sin(\varphi) \cos(\theta) y_n +  \sin(\theta) z_n) + \frac{y_n^2+z_n^2}{2d},
\end{align}
which is tight in near-broadside directions if $d$ is at least twice the aperture length \cite{Bjornson2021b}.
We rewrite the array gain using \eqref{eq:Taylor_approx}:
\begin{align} \nonumber
    G(F,d_{\textrm{r}}) &\approx \left| \sum_{n=1}^{N}
     e^{+j \frac{2\pi}{\lambda} \frac{y_n^2+z_n^2}{2F} } e^{j \epsilon_n}
     e^{-j \frac{2\pi}{\lambda} \frac{y_n^2+z_n^2}{2d_{\textrm{r}}} }
     \right|^2 \\
     & = \left| \sum_{n=1}^{N}
     e^{-j \frac{\pi}{\lambda} (y_n^2+z_n^2) \left( \frac{F-d_{\textrm{r}}}{d_{\textrm{r}}F} \right) } e^{j \epsilon_n} \right|^2. \label{eq:arraygain}
\end{align}
The focal point deviation in \eqref{eq:arraygain} can be represented by  a parameter $\delta$ defined as $\delta=\frac{1}{d_{\textrm{r}}}-\frac{1}{F} = \frac{F-d_{\textrm{r}}}{d_{\textrm{r}}F}$.
By treating $\epsilon_1,\ldots,\epsilon_N$ as independent random variables uniformly distributed in $\mathcal{Q}_b$, we can calculate the average array gain as
\begin{align} \nonumber
    &\bar{G}(F,d_{\textrm{r}}) \triangleq \mathbb{E} \{ G(F,d_{\textrm{r}}) \} \\
    &\approx 
    \sinc^2 \left( 2^{-b} \right)  \left| \sum_{n=1}^{N}
     e^{-j \frac{\pi}{\lambda}  (y_n^2+z_n^2)\delta } \right|^2 \!+ N \! \left(1 -\sinc^2 \!\left( 2^{-b} \right) \right) \label{eq:arraygain2}
\end{align}
following a similar approach as when we evaluated \eqref{eq:h_gain}.

To further simplify \eqref{eq:arraygain2}, we 
assume a uniform planar array (UPA) with $\sqrt{N} \times \sqrt{N}$ elements and a horizontal/vertical spacing of $\Delta$. 
The sum in \eqref{eq:arraygain2} cannot be computed in exact closed form, but since the element-spacing in an RIS is relatively small, we can treat it as the two-dimensional Riemann sum of the function $f(y,z)= e^{-j \frac{\pi}{\lambda} (y^2+z^2)\delta }$ over the square $\mathcal{S} = \{ (y,z) : -\Delta \sqrt{N}/2 \leq y,z \leq \Delta \sqrt{N}/2\}$ with interval length $\Delta$. Hence, it follows that
\begin{align} \nonumber
    \left| \sum_{n=1}^{N}
     e^{-j \frac{\pi}{\lambda}  (y_n^2+z_n^2)\delta } \right|^2 &\approx \left| \frac{1}{\Delta^2} \iint_{(y,z) \in \mathcal{S}} f(y,z) dy dz \right|^2 \\
    &  = \frac{1}{\Delta^4} \left| \int_{-\Delta \sqrt{N}/2}^{\Delta \sqrt{N}/2} e^{-j \frac{\pi}{\lambda} y^2 \delta } dy\right|^4, \label{eq:Riemann_sum}
\end{align}
where we recognized that the double integral can be divided into the product of two identical integrals.
The result can be evaluated using the Fresnel integrals\footnote{The Fresnel integrals are defined as $C(x) = \int_{0}^{x} \cos(\pi t^2/2) dt$ and $S(x) = \int_{0}^{x} \sin(\pi t^2/2) dt$. These can be evaluated using the error function.} $C(\cdot)$ and $S(\cdot)$ as
\begin{align} \label{eq:Fresnel_expression}
    N^2 \underbrace{\left( \frac{8 }{d_{\textrm{F}} \delta} \right)^{\! 2} \left( C^2 \left( \!\sqrt{\frac{d_{\textrm{F}} \delta}{8 } }\right) \!+\! S^2 \left( \!\sqrt{\frac{d_{\textrm{F}} \delta}{8 } }\right) \right)^2}_{\triangleq A(d_{\textrm{F}} \delta / 8)},
\end{align}
where $d_{\textrm{F}} = 4N \Delta^2/\lambda$ denotes the Fraunhofer distance computed for the UPA's diagonal aperture length $\sqrt{2N} \Delta$. 
The novel expression in \eqref{eq:Fresnel_expression} shows that the maximum gain, $N^2$, is multiplied by the function $A(x) = (C^2(\sqrt{x})+S^2(\sqrt{x}))^2/x^2$ with $x=d_{\textrm{F}} \delta/8$. This function previously appeared in studies of continuous aperture arrays, such as \cite{Polk1956a,Bjornson2021b}, and is a sinc-squared-like function with its maximum at $x=0$ ($d_{\textrm{r}}=F$). 

Hence, the average array gain when a $b$-bit RIS is defocused in the distance domain is
\begin{align} 
    \bar{G}(F,d_{\textrm{r}}) \approx 
    N^2 \sinc^2 \left( 2^{-b} \right) A \left( \frac{d_{\textrm{F}} \delta}{8} \right)+ N  \left(1 -\sinc^2 \!\left( 2^{-b} \right) \right). \label{eq:arraygain3}
\end{align}
Although there will not be any statistical averaging over the phase deviations $\epsilon_1,\ldots,\epsilon_N$ in practical LOS scenarios, we expect \eqref{eq:arraygain3} to be close to the true array gain whenever $N$ is large (following law of large numbers arguments).
We will validate this expression experimentally in Section~\ref{sec:experiments}.

When $N$ is large, the first term in \eqref{eq:arraygain3} dominates and the dependence on the bit-resolution $b$ and defocusing parameter $\delta$ decouples into separate factors.
This implies that a reduced bit resolution will lower the array gain by the same factor at all distances in the considered beam direction, but the shape of the beamfocusing effect is otherwise the same.

\subsection{Width of Beamfocusing}

The beamwidth around the focal point can be analyzed similarly to the depth.
Suppose the RIS is configured to focus its reflected signal on a point with the correct distance but incorrect angles.
Suppose the RIS configuration is selected as in \eqref{eq:phi_n_errors} but with $\varphi_{\textrm{r}},\theta_{\textrm{r}}$ replaced by $\varphi_{\textrm{F}},\theta_{\textrm{F}}$, which is a point on the curve in Fig.~\ref{focusing_example}.
The channel gain at the receiver then becomes
\begin{align} 
    |h|^2 = |\alpha_{\textrm{r}} \alpha_{\textrm{t}} |^2
    \underbrace{\left| \vect{a}^{\Htran}(\varphi_{\textrm{F}},\theta_{\textrm{F}},d_{\textrm{r}}) \diag (e^{j \epsilon_1},\ldots,e^{j \epsilon_N} ) \vect{a}(\varphi_{\textrm{r}},\theta_{\textrm{r}},d_{\textrm{r}}) \right|^2}_{\triangleq G(\varphi_{\textrm{r}},\theta_{\textrm{r}}, \varphi_{\textrm{F}},\theta_{\textrm{F}})}. \label{eq:h2_beamfocusing_width}
\end{align}
We can rewrite the array gain $G(\varphi_{\textrm{r}},\theta_{\textrm{r}}, \varphi_{\textrm{F}},\theta_{\textrm{F}})$
using \eqref{eq:Taylor_approx} as
\begin{align} 
    G(\varphi_{\textrm{r}},\theta_{\textrm{r}}, \varphi_{\textrm{F}},\theta_{\textrm{F}}) &\approx \left| \sum_{n=1}^{N}
     e^{-j \frac{2\pi}{\lambda} (B y_n +  C z_n) } e^{j \epsilon_n}
     \right|^2 , \label{eq:arraygain_width}
\end{align}
where we introduce the notation 
\begin{align}
\label{eq:B} 
    B &= \sin(\varphi_{\textrm{F}}) \cos(\theta_{\textrm{F}}) - \sin(\varphi_{\textrm{r}}) \cos(\theta_{\textrm{r}}) \\
    C &= \sin(\theta_{\textrm{F}}) - \sin(\theta_{\textrm{r}}) \label{eq:C}
\end{align} 
to capture the angle deviations. By once again treating $\epsilon_1,\ldots,\epsilon_N$ as independent random variables uniformly distributed in $\mathcal{Q}_b$, the average of the array gain in \eqref{eq:arraygain_width} becomes
\begin{align} \nonumber
    &\bar{G}(\varphi_{\textrm{r}},\theta_{\textrm{r}}, \varphi_{\textrm{F}},\theta_{\textrm{F}}) \triangleq \mathbb{E} \{ G(\varphi_{\textrm{r}},\theta_{\textrm{r}}, \varphi_{\textrm{F}},\theta_{\textrm{F}}) \} \\
    &\approx 
    \sinc^2 \left( 2^{-b} \right)  \left| \sum_{n=1}^{N}
     e^{-j \frac{2\pi}{\lambda}  (B y_n +  C z_n) } \right|^2 \!+ N \! \left(1 -\sinc^2 \!\left( 2^{-b} \right) \right)\!.
\end{align}
For a UPA with $\sqrt{N} \times \sqrt{N}$ elements with $\Delta$-spacing, we can approximate the sum as an integral (similarly to \eqref{eq:Riemann_sum}) to obtain
\begin{align} \nonumber
    \left| \sum_{n=1}^{N}
     e^{-j \frac{2\pi}{\lambda}  (B y_n +  C z_n) } \right|^2 
     & \! \approx \left| \frac{1}{\Delta^2} \! \iint_{(y,z) \in \mathcal{S}} e^{-j \frac{2\pi}{\lambda}  (B y +  C z) } dy dz \right|^2 \\
     & \!\!\!\!\!\!\!\!\!\!\!\!\!\! = N^2 \sinc^2 \left( \frac{B \Delta \sqrt{N}}{ \lambda} \right) \sinc^2 \left( \frac{C \Delta \sqrt{N}}{ \lambda} \right) \!,
\end{align}
where we utilize the fact that $\int_{-L}^{L} e^{j \pi D x} dx = 2L \sinc(DL)$ for arbitrary real constants $L$ and $D$.
Hence, the average array gain when a $b$-bit RIS is defocused in the angular domain is
\begin{align} \nonumber
    &\bar{G}(\varphi_{\textrm{r}},\theta_{\textrm{r}}, \varphi_{\textrm{F}},\theta_{\textrm{F}}) \approx 
N  \left(1 -\sinc^2 \!\left( 2^{-b} \right) \right) \\
&+
N^2 \sinc^2 \left( 2^{-b} \right) \sinc^2 \left( \frac{B \Delta \sqrt{N}}{ \lambda} \right) \sinc^2 \left( \frac{C \Delta \sqrt{N}}{ \lambda} \right). \label{eq:arraygain3_width}
\end{align}
The dependence on the angle mismatches appears through $B$ and $C$, which are defined in \eqref{eq:B}--\eqref{eq:C}. This expression has a sinc-squared behavior, which is relatively common in beamwidth analysis involving UPAs (cf.~\cite[Eq.~(32)]{Bjornson2021b}). Interestingly, when $N$ is large, the bit-resolution does not affect the beamwidth, which is determined by the nulls of the last two sinc-squared functions in \eqref{eq:arraygain3_width}. 
Hence, even if the hardware resolution limits the maximum array gain, it does not change the shape of the beam.
We will validate this expression experimentally in the next section.

\section{Experimental Validation} \label{sec:experiments}
In this section, we present experimental results that validate the ability of performing dynamic beamfocusing using an RIS, and compare the results to the theoretical formulas derived in Section~\ref{sec:theoretical_formulas} regarding the beamdepth and beamwidth.

\begin{figure} 
        \centering 
        \begin{subfigure}[b]{\columnwidth} \centering 
	\begin{overpic}[width=\columnwidth,tics=10]{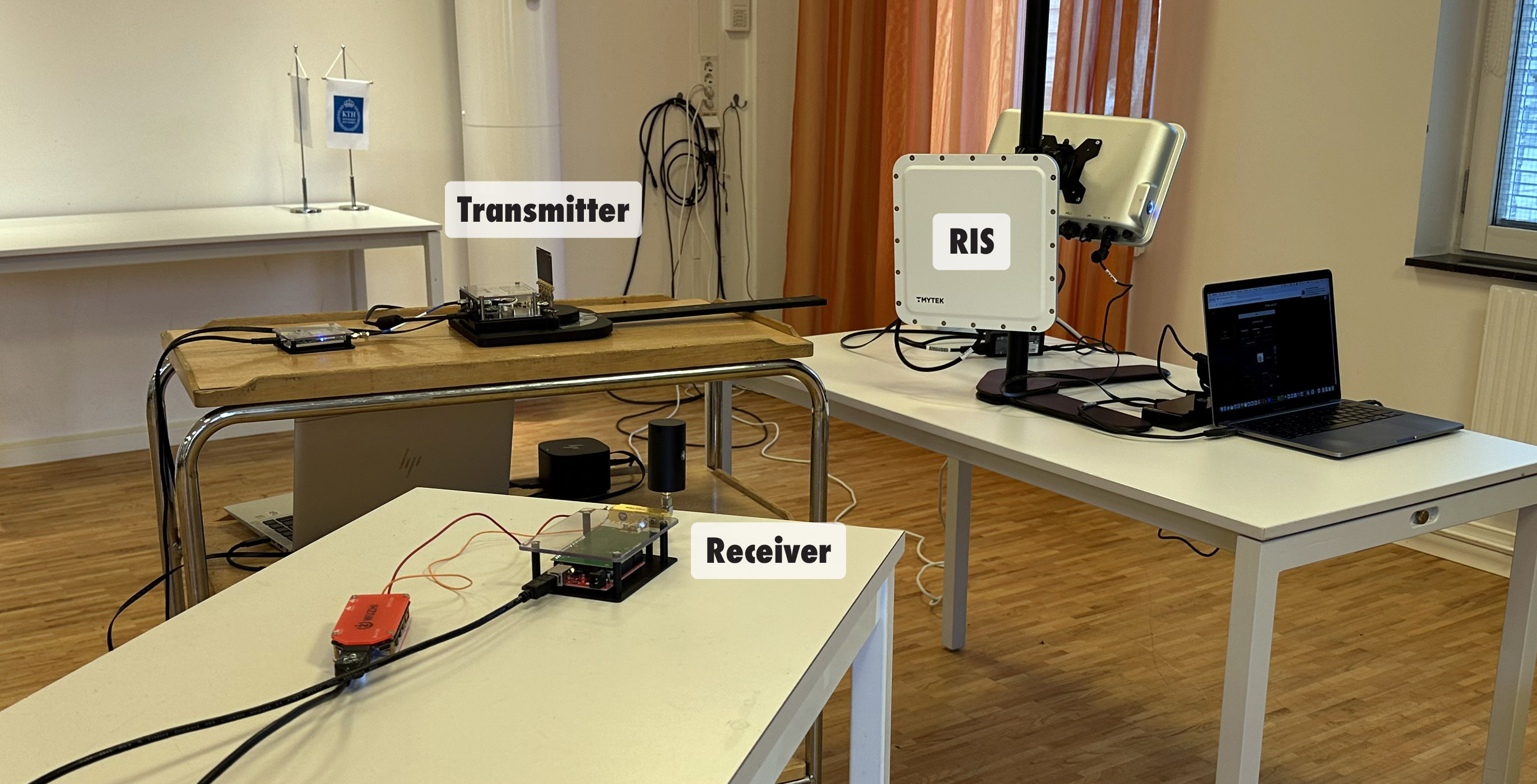}
 \end{overpic}    
                \caption{The first measurement setup.}   \vspace{+2mm}
        \end{subfigure} 
        \begin{subfigure}[b]{\columnwidth} \centering
	\begin{overpic}[width=\columnwidth,tics=10]{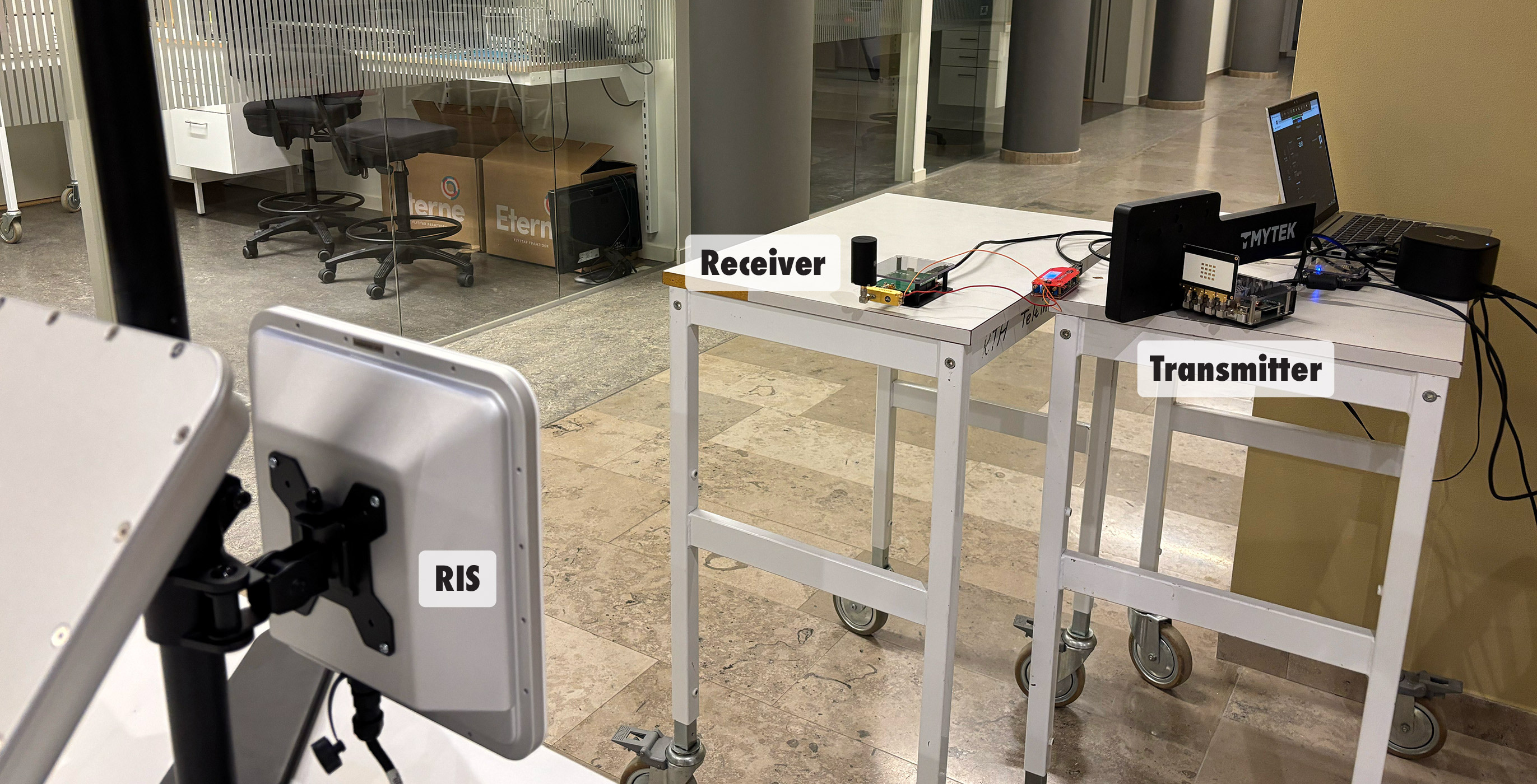}
\end{overpic} 
                \caption{The second measurement setup.}  
        \end{subfigure} 
        \caption{The setups used for experimental validation of reflective near-field beamfocusing from a dynamic RIS. The measurements were made at the KTH Royal Institute of Technology.} \vspace{-2mm}
        \label{ExpSetup}  
\end{figure}

The testbed is shown in Fig.~\ref{ExpSetup}, operates in the mmWave band around 28 GHz, and consists of three main components from TMYTEK.\footnote{These components come from the 5G mmWave Developer Kit and XRifle Dynamic 28 GHz RIS (2024 editions) from TMYTEK, \url{https://tmytek.com/}}
The first component is the transmitter with $4 \times4$ radiating elements in UPA geometry. It was rotated so the RIS is in its boresight and all elements were fed with the same signal to create a boresight beam toward the RIS. Hence, it acted as a single-antenna transmitter within our experiments, as assumed in the theoretical derivations.
The second component is the RIS, which consists of $N=1024$ reflecting elements, arranged as a $32 \times 32$ UPA with half-wavelength spacing. Each element is equipped with a $1$-bit phase shifter that can implement two different phase-shifts separated by $\pi$, i.e., $\phi_{n} \in \{0, \pi\}$. Hence, the phase deviations from the optimal values belong to $\mathcal{Q}_1=[-\frac{\pi}{2},\frac{\pi}{2}]$.
The third component is the receiver, equipped with an antenna having an omnidirectional azimuth pattern and a power detector that registers the received power. All the equipment was deployed at the same height above the ground, so only azimuth angles and distances vary across the measurements.

The phase configuration $\boldsymbol{\Phi} = \diag\left(e^{j\phi_{1}}, \ldots, e^{j\phi_{N}}\right)$ of the RIS is determined by the TMYTEK software, which takes angles and distances to the transmitter and receiver (with respect to the RIS) as input. The exact calculation is not disclosed in the user manual, but we believe it aligns the phases similarly to 
\eqref{eq:phi_n} with $h_{\textrm{s}} = 0$ and quantizes the results to the nearest implementable value. The software compensates for mutual coupling and other RIS hardware impairments.

The Fraunhofer distance of the RIS is $d_{\textrm{F}} = 10.8$\,m, which covers the entire room used in the first measurement setup. However, the depth of focus depends on where the receiver is located.
To demonstrate this experimentally, we will keep the transmitter and receiver at fixed locations, and vary the focal distance $F$ that is used when configuring the RIS.\footnote{The depth properties of a beam are normally plotted by keeping the focal distance $F$ fixed and varying the receiver distance $d_{\textrm{r}}$. However, we utilized the fact that the channel gain in \eqref{eq:h2_beamfocusing} is symmetric in $F$ and $d_{\textrm{r}}$ to do the opposite, as it is challenging to move mmWave equipment with high precision.}

\subsection{First Measurement Setup}

The first measurement setup is shown in Fig.~\ref{ExpSetup}(a) and the goal was to validate the beamfocusing behavior in the depth domain.
The transmitter is located at $\varphi_{\textrm{t}}=-25^\circ$, $\theta_{\textrm{t}}=0^\circ$, and $d_{\textrm{t}}=1$\,m. The receiver is located at $\varphi_{\textrm{r}}=7^\circ$ and $\theta_{\textrm{r}}=0^\circ$ (these parameters are fixed in the RIS configuration), and the distance $d_{\textrm{r}}$ is varied across measurements.
We ensured there is no noticeable static path in this setup (i.e., $h_{\textrm{s}}\approx 0$).

\begin{figure}
    \centering
    {\includegraphics[width=\columnwidth]{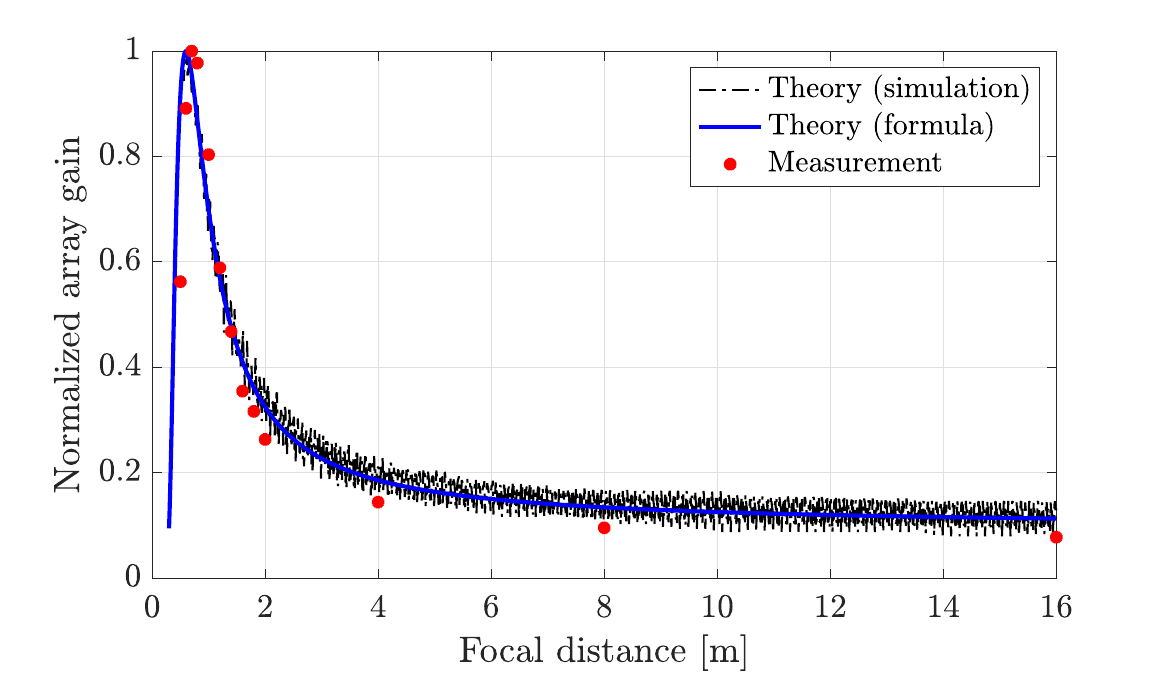}}\hfill
    \caption{The array gain in Setup 1 with $d_{\textrm{r}}=0.6$\,m.}
    \label{fig:1} \vspace{-3mm}
\end{figure}

In Fig.~\ref{fig:1}, we show the array gain (normalized to have the peak value $1$) as a function of the focal distance $F$ when the receiver distance is $d_{\textrm{r}}=0.6$\,m.
We compare the measured values with the theoretical formula in \eqref{eq:arraygain3} and with a simulation where the phase-shifts were computed using \eqref{eq:phi_n} and then quantized to the nearest points in $\mathcal{Q}_1$.
The measured and theoretical curves show good agreement, with both exhibiting a sharp focusing effect around $F=d_{\textrm{r}}$. This demonstrates that our new formula properly captures how the array gain depends on the distance under 1-bit quantization. The half-power beamdepth is around $1$\,m.
The dash-dotted black simulation curve fluctuates rapidly when the RIS is defocused, due to the quantization effect in the 1-bit phase-shifts. The solid blue curve, based on the formula in \eqref{eq:arraygain3}, appears at the center of these fluctuations. This validates that this formula accurately approximates the average array gain.

\begin{figure}
    \centering 
    {\includegraphics[width=\columnwidth]{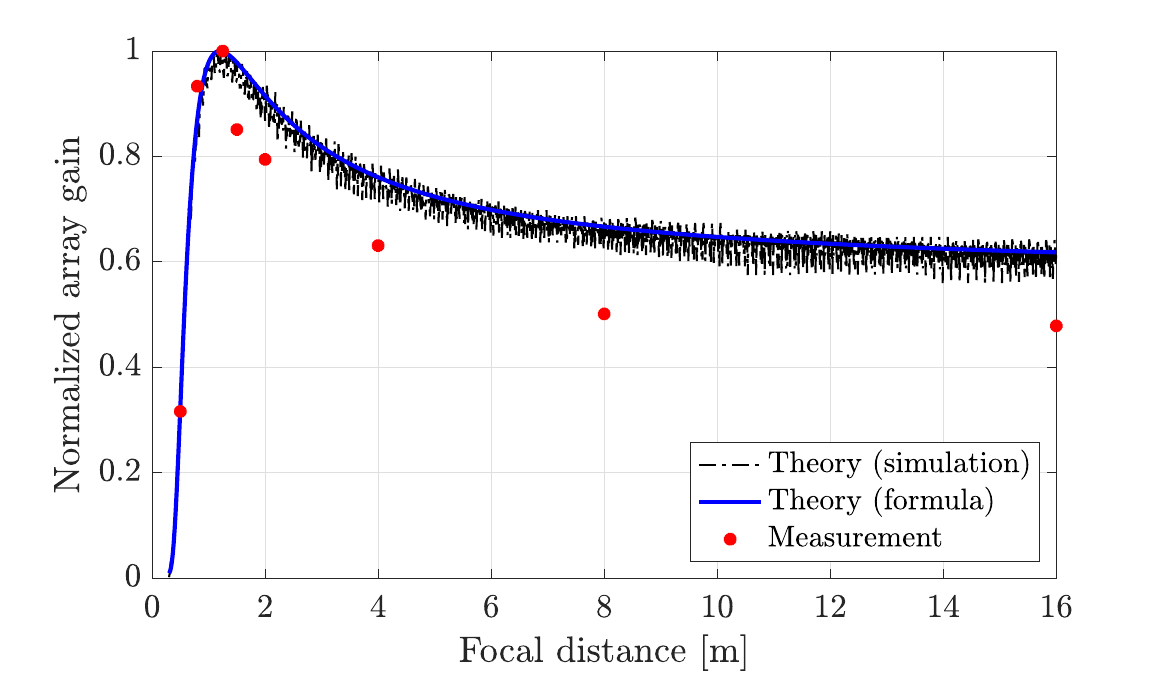}}\hfill
    \caption{The array gain in Setup 1 with $d_{\textrm{r}}=1.2$\,m.}
    \label{fig:2}
\end{figure}

In Fig.~\ref{fig:2}, we increase the receiver distance to $d_{\textrm{r}}=1.2$\,m, while keeping all other parameters constant. The theoretical and experimental results maintain consistent focusing characteristics, and the measured peak aligns closely with the theoretical focus. Interestingly, the measured curve exhibits a slightly narrower beamdepth than the theoretical curves. This indicates that practical characteristics not considered in our system model (e.g., mutual coupling, gain variations across elements, and additional propagation paths) will not destroy the beamfocusing ability of the RIS, but rather strengthen it.

\subsection{Second Measurement Setup}

The second measurement setup is shown in Fig.~\ref{ExpSetup}(b) and the goal was to jointly demonstrate beamfocusing in both angle and depth.
The transmitter is located at $\varphi_{\textrm{t}}=-17^\circ$, $\theta_{\textrm{t}}=0^\circ$, and $d_{\textrm{t}}=1$\,m. The receiver is located at  $\varphi_{\textrm{r}}=\theta_{\textrm{r}}=0^\circ$ with the distance $d_{\textrm{r}}=0.8$\,m.
We ensured there was no noticeable static path by placing a metal object between the transmitter and receiver, as this is assumed in the theoretical formulas.

\begin{figure} 
        \centering 
        \begin{subfigure}[b]{\columnwidth} \centering 
	\begin{overpic}[width=\columnwidth,tics=10]{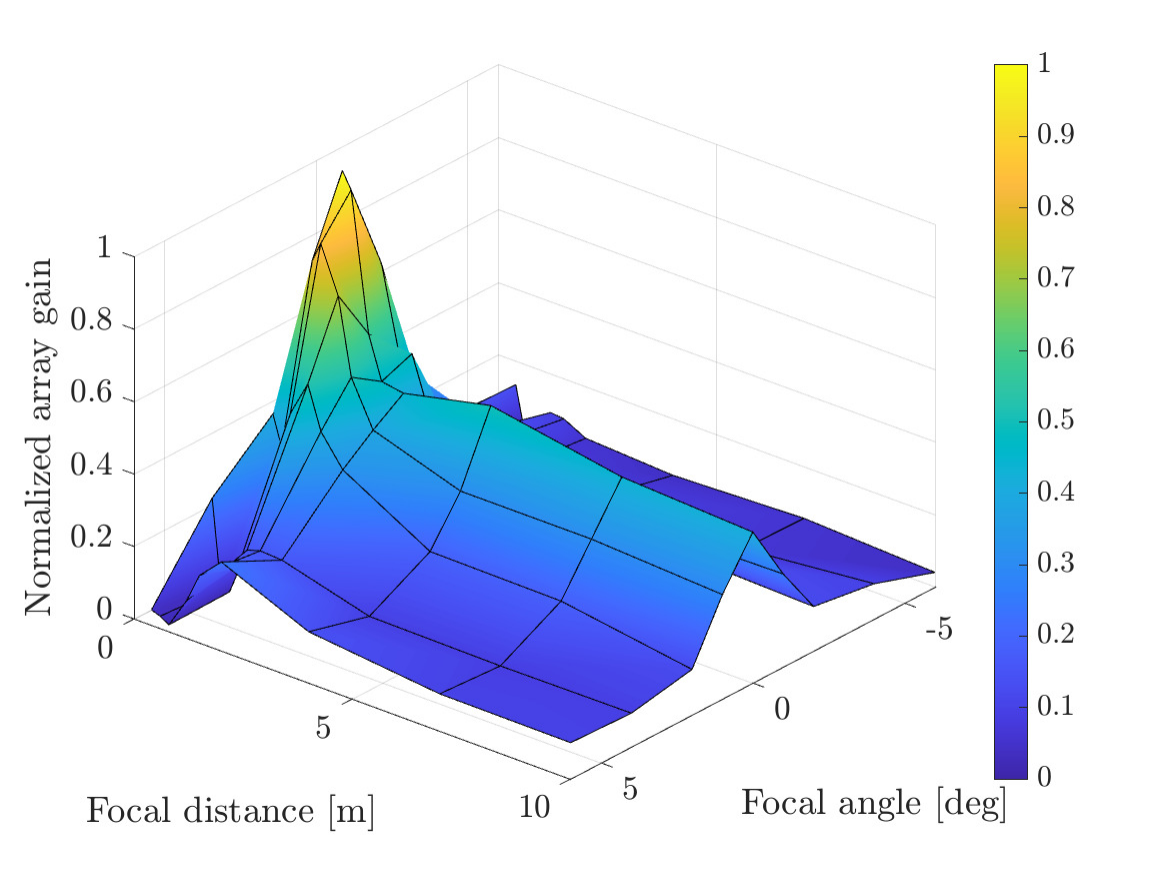}
 \end{overpic}    
                \caption{Three-dimensional plot.}   \vspace{+2mm}
        \end{subfigure} 
        \begin{subfigure}[b]{\columnwidth} \centering
	\begin{overpic}[width=\columnwidth,tics=10]{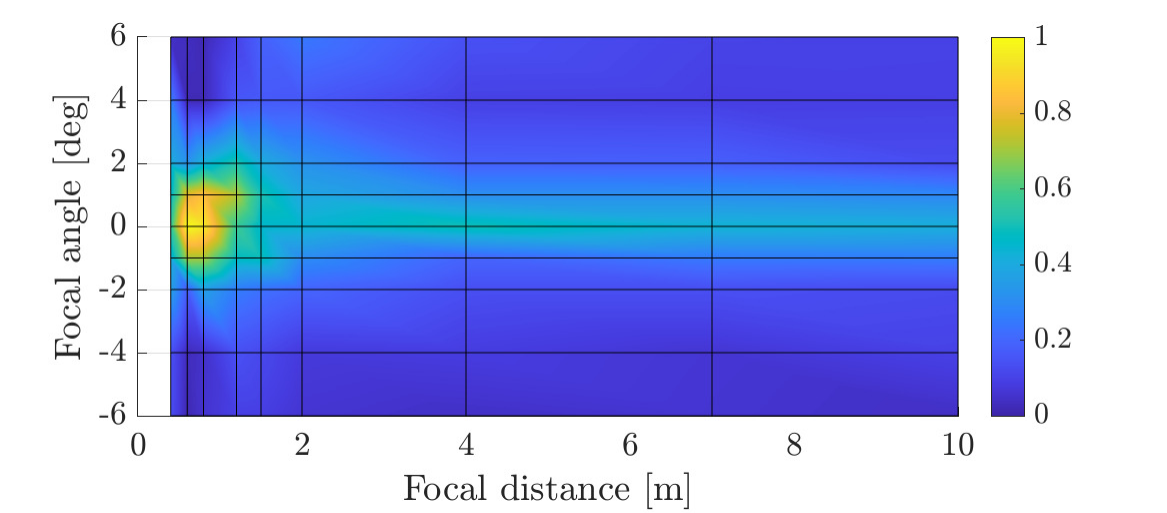}
\end{overpic} 
                \caption{Same plot but viewed from above.}  
        \end{subfigure} 
    \caption{The array gain in Setup 2 when varying both the focal distance and azimuth angle around the receiver.}
    \label{fig:3}
\end{figure}

Fig.~\ref{fig:3} shows how the array gain varies when the focal point is moved in both distance and azimuth angle, as compared to the boresight receiver at $d_{\textrm{r}}=0.8$\,m. This corresponds to moving along the two axes shown in Fig.~\ref{focusing_example}.
All the intersection points in this graph represent the measurements, while the surface and color indicate the shape. The yellow peak value coincides with the true location of the receiver, and there is a roughly ellipsoidal area around it with a substantial array gain (shown in yellow and green). This result demonstrates that one can achieve beamfocusing with a textbook-like shape, as sketched in Fig.~\ref{focusing_example}, in a real office environment.

\begin{figure}
    \centering
    {\includegraphics[width=\columnwidth]{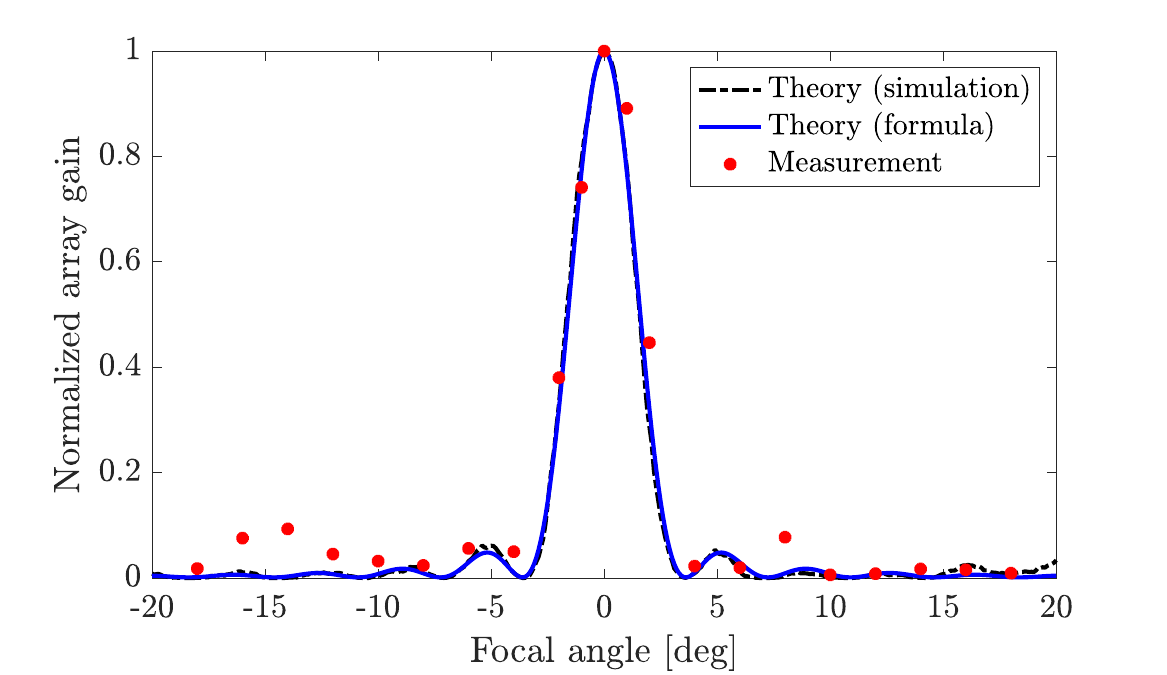}}\hfill
    \caption{The array gain in Setup 2 when varying the focal angle in the azimuth plane.} \vspace{-3mm}
    \label{fig:4}
\end{figure}

Next, we take a closer look at the beamwidth behavior by varying the azimuth angle of the focal point in the range $[-20^\circ,20^\circ]$, while keeping the focal distance and elevation angle the same as for the receiver.
Fig.~\ref{fig:4} shows the measurement results and two theoretical curves. The solid blue curve is computed using the formula in \eqref{eq:arraygain3_width} and matches the measurements closely, particularly within the mainlobe. The half-power beamwidth is around $3$ degrees.
The black dash-dotted curve is once again computed by calculating theoretical LOS channels and quantizing the ideal RIS phase-shifts. This simulation curve is close to the formula-based curve, but when it deviates in the sidelobes, it is slightly closer to the measurements.
There is a minor discrepancy between measurements and theory around $-15^\circ$. In this case, the RIS reflects the signal back toward the transmitter, and the metal object that we used to block the direct path between the transmitter and receiver reflects this signal toward the receiver.

\section{Conclusions}
\label{sec:conclusion}

The near-field communication literature frequently highlights finite-depth beamfocusing as an attractive new interference-suppression feature that concentrates radiated signals toward a spatial region confined in both angle and depth.
In this paper, we have demonstrated that this theoretically appealing concept can be realized in practical indoor environments using a dynamic $b$-bit RIS.
We derived novel analytical expressions for the depth and width of the near-field reflective beamfocusing effect and validated them experimentally in 28 GHz measurement setups, considering two different office environments.
The close agreement between theoretical predictions and measured results confirms that the derived formulas accurately capture real-world behaviors, even if derived from a system model that neglects mutual coupling, element gain variations, and multipath propagation. 
Importantly, this indicates that more advanced electromagnetic models are not needed when studying LOS setups where the RIS hardware has been designed to actively compensate for hardware impairments.
To the best of our knowledge, this paper provides the first experimental validation of reflective near-field beamfocusing with a dynamic RIS, establishing its feasibility for future RIS-assisted communication systems.

\section*{Acknowledgment}

The authors would like to thank 
Elina Björnson,
Anders Enqvist, and
Eren Berk Kama for supporting the experiments.

\bibliographystyle{IEEEtran}
\bibliography{IEEEabrv,refs}

\end{document}